\documentclass[aps,preprint,onecolumn,groupedaddress]{revtex4}

\usepackage{graphicx}
\usepackage{hyperref}
\usepackage{epsfig}
\begin{document} 

% \DeclareGraphicsExtensions{.eps,.EPS,.jpg,.bmp}

%\preprint{}

\title{Long-range temperature-controlled transport of ultra-cold atoms with an accelerated lattice}


\author{Luc Absil}
%\email{luc@absil.fr}
\author{Yann Balland}
%\email{yann.balland@obspm.fr}
\author{Franck Pereira dos Santos}
\email{franck.pereira@obspm.fr}

% \email{}
%\thanks{}
\affiliation{LNE-SYRTE, Observatoire de Paris, Université PSL,
CNRS, Sorbonne Université, Paris, F–75014, France}


\date{\today}

\begin{abstract}
We report our method for transporting ultracold atoms over macroscopic distances and trapping them back in a vertical mixed trap, consisting of the superposition of a vertical lattice and a transverse confinement beam. The transport is performed with Bloch oscillations allowing us to move up to $25\%$ of a sub-micro-Kelvin atomic cloud on a distance of the order of $30$cm, without excessive heating and with a good control of its final position. The efficiency is lowered to about $10\%$ after trapping them back in the vertical mixed trap during extended times.

\end{abstract}

 %\keywords{}

\maketitle

\section{Introduction}

The transport of clouds of cold atoms or Bose-Einstein condensates (BEC) from their production area to other areas of experimental interest has drawn much attention since first implementations in the 90s \cite{peikBlochOscillationsAccelerator1997}. 
Various approaches have been considered, using magnetic or optical fields, depending on the desired specifications. 
Magnetic traps have demonstrated to be a suitable solution for long-range displacement of tens of centimeters, 
but at the cost of heating up the sample \cite{greinerMagneticTransportTrapped2001}. 
On the other hand, trapping the atomic cloud in optical lattices has proven to be a solution of choice
for precise displacement of single atoms over distances of the order of the centimeter \cite{kuhrDeterministicDeliverySingle2001} \cite{schraderOpticalConveyorBelt2001}, for adiabatic centimeter-long transport of ultracold atoms \cite{middelmannLongrangeTransportUltracold2012}. Other solutions involving the use of Bessel beams \cite{klostermannFastLongdistanceTransport2022}, of a focus-tunable moiré lens \cite{unnikrishnanLongDistanceOptical2021}, other optical systems of tunable-zoom \cite{leeTransportingColdAtoms2020} or optical tweezers \cite{couvertOptimalTransportUltracold2008} have also shown their interest to transport atomic cloud either on short distances with reasonably small heating or on long distances with more consequent warm-up of the atomic cloud.  
 
In the experiment reported here, we aim at performing ultra-sensitive short range forces measurements, with a quantum sensor based on $\mathrm{^{87}Rb}$ atoms trapped in an optical trap, consisting of a vertical shallow lattice $\lambda_{Verdi}=532$ nm overlapped with a transverse confinement potential $\lambda_{IR} = 1064$ nm. In such a system, energy differences between adjacent lattice sites can be probed with a Ramsey-Raman interferometer \cite{alauzeTrappedUltracoldAtom2018} providing us a local measurement of the vertical force exerted on the atoms. The goal is to probe micrometer-range forces occurring between our atomic cloud and the surface of a mirror.
Previous attempts at similar measures have shown the importance of preparing the sample away from the surface of interest in order to minimize any undesired adsorption of the atoms onto the surface
\cite{sorrentinoQuantumSensorAtomsurface2009a} 
\cite{mcguirkAlkaliAdsorbatePolarization2004}. Furthermore, the spatial resolution of the measurements is limited by the final size of the cloud, in the first place in the vertical direction, perpendicular to the surface of the mirror, of the order of a few micrometers \cite{alauzeTrappedUltracoldAtom2018}. Secondly, the transverse elongation of the cloud leads to an averaging of the roughness of the surface. 

~\\
We present here the outcomes of our experimental setup which enables the transportation of up to $25\%$ of a sub-micro-Kelvin atomic cloud on a distance of the order of 30 cm, without inducing excessive heating and while maintaining control over its size and final position. We first provide a general overview of the experimental setup and of the main sources of loss expected, before reporting our results on the transport efficiency and the control of the atoms position.

%%%%%%%%%%%%%%%%%%%%%%%%%%%%%%%%%%%%%%%%%%%%%%%%%%%%%%%
%%%%%%%%%%%%%%%%%%%%%%%%%%%%%%%%%%%%%%%%%%%%%%%%%%%%%%%
\section{Experimental setup}

\begin{figure}[h]
    \begin{center}
      \includegraphics[scale=.8]{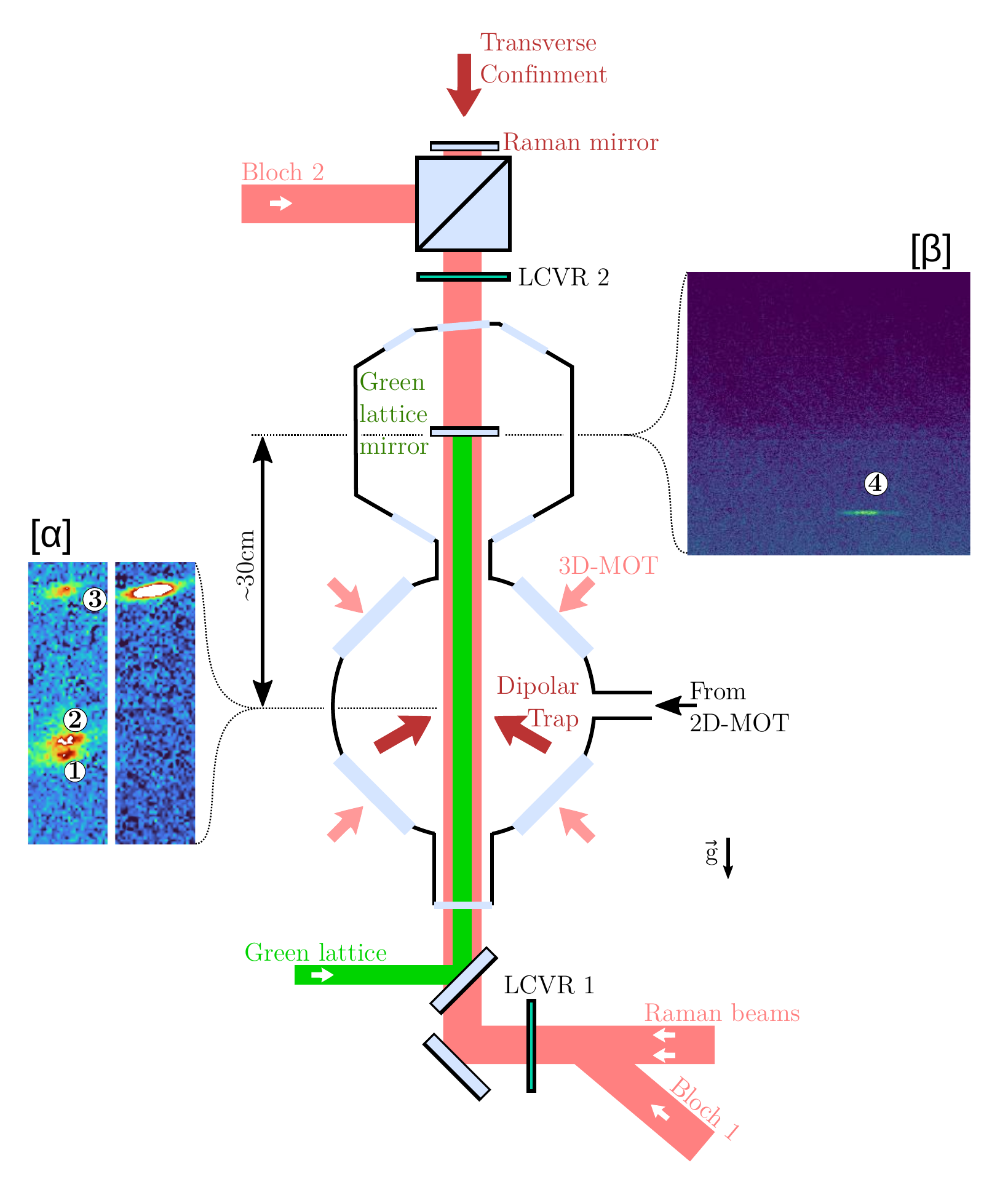}
    \end{center}
  \caption{Experimental setup of our experiment. \\ Subfigures $[\alpha]$ and $[\beta]$ provide examples of absorption images of the atomic cloud, respectively 3ms after the starting of the elevator and 30cm higher, at the end of the elevator, near the surface of interest. In $[\alpha]$ two situations are illustrated, either in the case of an imperfect (\textit{left}) polarization, or in the case of an adjusted (\textit{right}) polarization of the ``\textit{Bloch 1}`` beam, which emphasizes the losses due to the parasitic lattice due to the reflection of the ``\textit{Bloch 1}`` beam on the Raman mirror (\raisebox{.5pt}{\textcircled{\raisebox{-.9pt} {2}}}) and the losses due to the inadiabaticity of the launch (\raisebox{.5pt}{\textcircled{\raisebox{-.9pt} {1}}}).}
  \label{VerticalSetup}
\end{figure}

\begin{figure}[h]
  \begin{center}
    \includegraphics[scale=1]{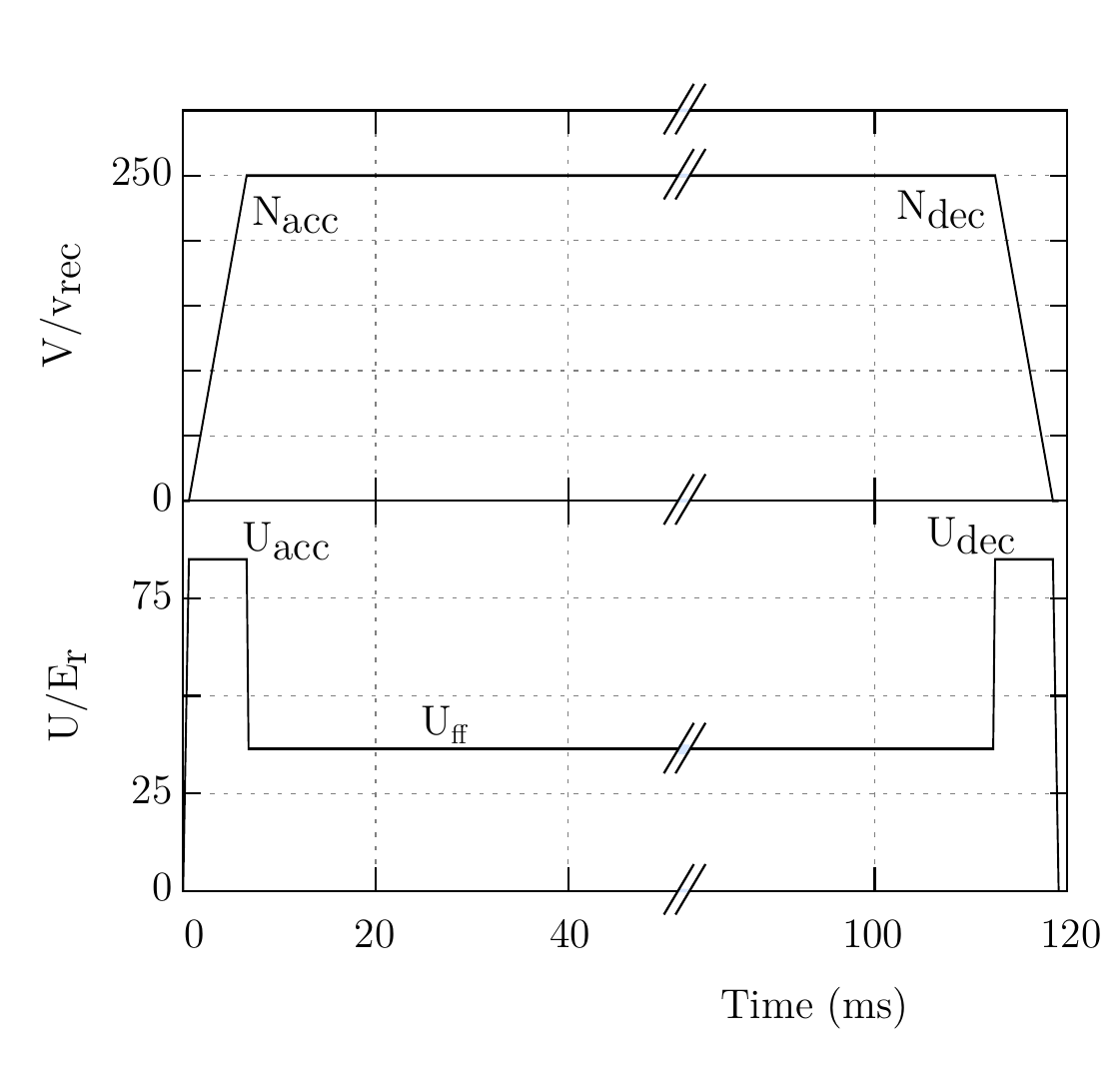}
  \end{center}
\caption{Optimized sequence for the laser power and frequency during the elevator phase.\label{BlochSetup}}
\label{Rampes}
\end{figure}

Figure \ref{VerticalSetup}-a summarizes the main steps of the atom preparation. A two-dimensional magneto-optical-trap (2D-MOT) loads atoms in a three-dimensional magneto-optical trap (3D-MOT), which in turn loads a crossed dipole trap where evaporative cooling is performed down to a temperature of 300 nK and an atomic density of the order of $10^{12}$ atoms$/$cm$^3$. Then, atoms are launched upward and transported over 30 cm, up to the vicinity of the surface of the lattice mirror, where measurements are to be performed.

The elevator consists of the overlapping of two counterpropagating beams ``\textit{Bloch 1}`` and ``\textit{Bloch 2}`` whose frequency differences and intensities are tuned over time. Atoms are thus loaded in an accelerated or decelerated lattice of adjustable depth. 

\subsection{Setup}

The Bloch beams are derived from a single external-cavity diode laser, which is split in two. Each beam is sent to a double pass acousto-optic modulator (AOM) before injecting a 2W tapered amplifier. Each amplified beam then passes through an other single pass AOM and then injects a polarization maintaining fiber. One fiber directs the first Bloch beam to the bottom of the experimental setup, where it enters the vacuum chamber, while the other fiber bring the second Bloch beam to the top. They are finally overlapped so as to counterpropagate along the vertical direction. This setup allows for modulating the frequency difference between the beams during the transport with the AOMs before the TAs, and their powers with the AOMs placed after the TAs.

\subsection{Atoms in moving lattices}

The periodic potential $U(z)$ resulting from the interference of the two counter-propagating beams ``\textit{Bloch 1,2}'' of intensity $I_{1,2}$ is proportional to the local intensity experienced by the atoms $I(z) = 2 I_0 (1 + m \cos 2kz)$, where $I_0$ is the average intensity and $m$ the ratio of the geometric and arithmetic means of the intensities of each beam :

\begin{equation}
  U(z) = \frac{U_0}{2} (1 + m \cos 2 k z)
\end{equation}

For a detuning $\Delta$ of the order of hundreds of GHz with respect to the D2 line of $^{87}$Rb, the depth is $U_0 \sim \hbar \Gamma^2 I_0 / (3 \Delta I_s)$, where $\Gamma$ and $I_s$ are the natural linewidth and the saturation intensity of the transition, and $k$ is the modulus of the wavevector of the laser. By imposing a frequency difference $\Delta\omega$ between the two Bloch beams, we then move the lattice potential with a velocity $v_{latt} = \Delta\omega/2k$. This allows us to coherently accelerate the atoms through Bloch oscillations \cite{bendahanBlochOscillationsAtoms1996}, each oscillation imparting a $2 \hbar k$ momentum transfer to the atoms.

For the measurements presented here, the detuning $\Delta$ was set to -250 GHz. For maximum powers of order of 50 mW in the two beams, with identical waists of 400 $\mu$m, we calculate a maximum depth of the order of $150 E_r$. This is in reasonable agreement with a determination of the depth based on Raman-Nath diffraction, where from the measurement of the populations in the diffracted orders as a function of duration of a square pulse performed before the transport, we deduced a maximum depth of about $110 E_r$.

Figure \ref{BlochSetup} illustrates the typical sequence of transport. The atoms are first adiabatically loaded into the lattice, by increasing the depth from 0 to $U_{acc}$ in 600 $\mu$s while ramping the frequency difference between the two beams by 25 MHz/s. This corresponds to accelerating the lattice by $g$. 

Then, the lattice is accelerated upwards over 6 ms at a fixed lattice depth $U_{acc}$. The chirp of the frequency difference between the beams corresponds to $N_{acc}$ Bloch oscillations. After a certain transport time at constant velocity, where the lattice potential may be reduced, the atoms are decelerated at a depth $U_{dec}$ with $N_{dec}$ Bloch oscillations. The atoms are then adiabatically released by reducing the lattice potential down to zero in 600 $\mu$s. They are then trapped in the mixed trap consisting of the superposition of the green lattice and the IR transverse confinement beam illustrated in figure \ref{VerticalSetup}, where force measurements are to be performed.

Between the acceleration and deceleration, we keep the lattice at a depth $U_{ff}$  in order to keep the atoms vertically and transversally confined. Additional transverse confinement is also provided by the IR laser at 1064 nm whose waist, of the order of 200 $\mu$m is placed at the target position of the atoms, near the surface of the lattice mirror.

\subsection{Sources of loss}
Common sources of loss during the transport with Bloch oscillations are spontaneous emission and non-adiabatic losses. Spontaneous emission occurs at a rate of order of $P_{sp} \sim U_0 \Gamma / \hbar \Delta$ and leads to losses that increase with the duration of the transport. Note that in our case, the tapered amplifiers emit in addition amplified spontaneous emission near resonance, which we efficiently filter using narrow bandwidth (of order of $\sim 150$ GHz) interference laser line filters. This in turn forces us to use large enough detunings.

For their efficient transport with Bloch oscillations, atoms have to be initially loaded and then to remain in the lower band of the moving lattice. The adiabaticity of the acceleration process restricts accelerations to below a critical value, which, in the weak binding limit, can be deduced from the Landau-Zener formula \cite{bendahanBlochOscillationsAtoms1996}. % Yann: Citation de la formule de LZ ? Citation de l'article original tant qu'à faire ?
The larger the depth, the larger this critical acceleration, and generally losses decrease with decreasing accelerations, and thus longer transport durations, by contrast with losses by spontaneous emission. This calls for a compromize between these two sources of losses. 
Note that for depths below 20 $E_r$, resonances between the energy bands also reduces the efficiency in certain acceleration ranges.

Finally, specific to our optical setup, a last source of loss is due to small fractions of the \textit{Bloch 1} beam that retroreflect on the green lattice and Raman mirrors. These create additional static lattices, in which a fraction of the atoms, can be trapped  as displayed in subfigure \ref{VerticalSetup}-$\alpha$. To minimize them, on one hand, the dichroic character of the green lattice mirror yields a low reflectivity of about $0.6\%$ at 780 nm, and on the other hand, the polarization beam splitter placed on the top should ideally prevents the Bloch 1 beam from reflecting onto the Raman mirror. In practice, a residual reflection is minimized by carefully adjusting the polarization of the beam. 

\subsection{Detection}
Two different detection schemes are used to evaluate the performance of our transport system. We use in-situ absorption imaging with a PCO PixelFly CCD camera in the bottom chamber, and both absorption imaging or fluorescence detection with a deeply cooled EM-CCD camera in the top chamber. This allows us to measure the number of atoms in the $| F=2 \rangle$ state and their temperature in both chambers. In-situ fluorescence imaging in the top chamber allows to measure with a better atom detectivity but without much spatial resolution the number of atoms in each of the $| F=1 \rangle$ and $| F=2 \rangle$ states. 
 
\section{Results}

To optimize the launch efficiency of the atoms, we started by imaging in the bottom chamber their position 3 ms after their acceleration. Subfigure \ref{VerticalSetup}-$\alpha$ left displays an absorption imaging picture featuring three different groups of atoms: (3) launched, (2) trapped in a parasitic lattice due to reflection of the "Bloch 1" beam on the Raman mirror, and (1) untrapped and in free fall. By adjusting the polarization of the ``Bloch 1`` beam, we eliminated the loss of atoms in both groups (1) and (2), which results in a single group of atoms as displayed at the right. We then increased the number of Bloch oscillations until the atoms could reach the top chamber, where we obtained a first signal (similar to the one in subfigure \ref{VerticalSetup}-$\beta$). We then optimized the durations of the different transport phases, keeping the transport distance constant. This initial optimization resulted in durations of 6 ms for both acceleration and deceleration phases, with 250 Bloch oscillations at the acceleration as displayed in figure \ref{Rampes}. We then optimized the other parameters, such as the power of the ``Bloch`` beams or the temperature of the atoms before the transport. We finally varied back the duration and number of Bloch oscillations without much improvement.

\begin{figure}[h]
    \begin{center}
      \includegraphics[scale=1]{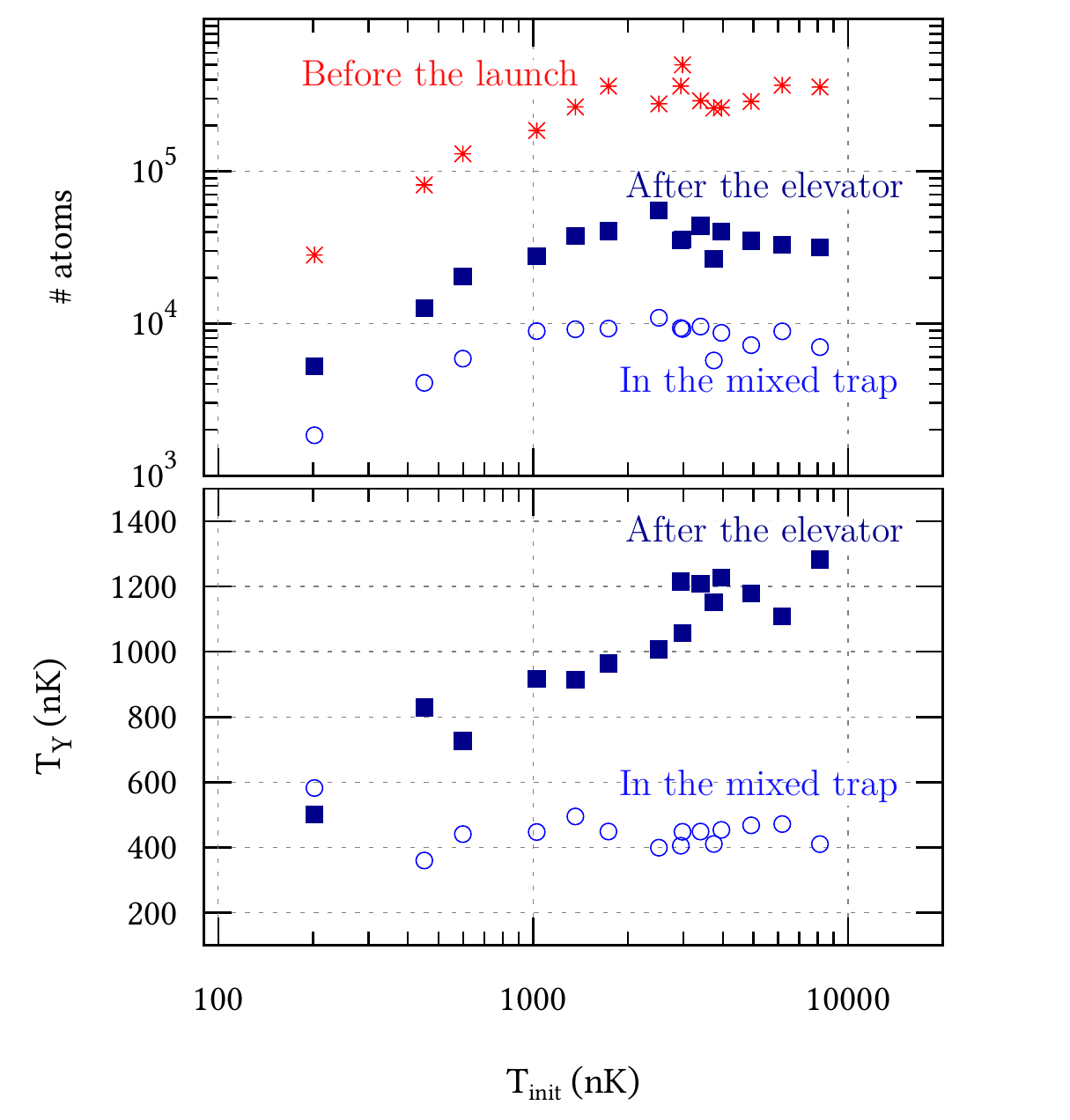} 
    \end{center}
  \caption{Influence of the initial temperature of the cloud (at the end of the cooling processes and right before the launch) on (top) the number of atoms at the end of the transport and after 1s in the mixed trap and (bottom) the temperature in the vertical direction.}
  \label{Tinit}
  \end{figure} 

We illustrate Fig. \ref{Tinit} the influence of the initial temperature of the atoms both on the number and the temperature of the atoms after their transport over the 30cm distance and after their trapping in the mixed trap described above. We find that the efficiency of both processes reduces when increasing the initial temperature. We attribute this for one to the velocity selectivity of the acceleration process, that works best for atoms moving at subrecoil velocities in the frame of the lattice, and for the other to the shallow depth of the mixed trap. Note that the efficiency is not the only relevant parameter when it comes to selecting the most appropriate parameters for our transport system. It can be of interest for instance to work either with high initial temperatures, which, despite the lower efficiency, leads indeed to larger number of transported atoms, or with lower initial temperature, which leads to smaller sizes at the end. These relevant parameters actually depend on the details of the evaporative cooling stage, where temperature, number of atoms and initial size are correlated.

\begin{figure}[h]
  \begin{center}
    \includegraphics[width=\textwidth]{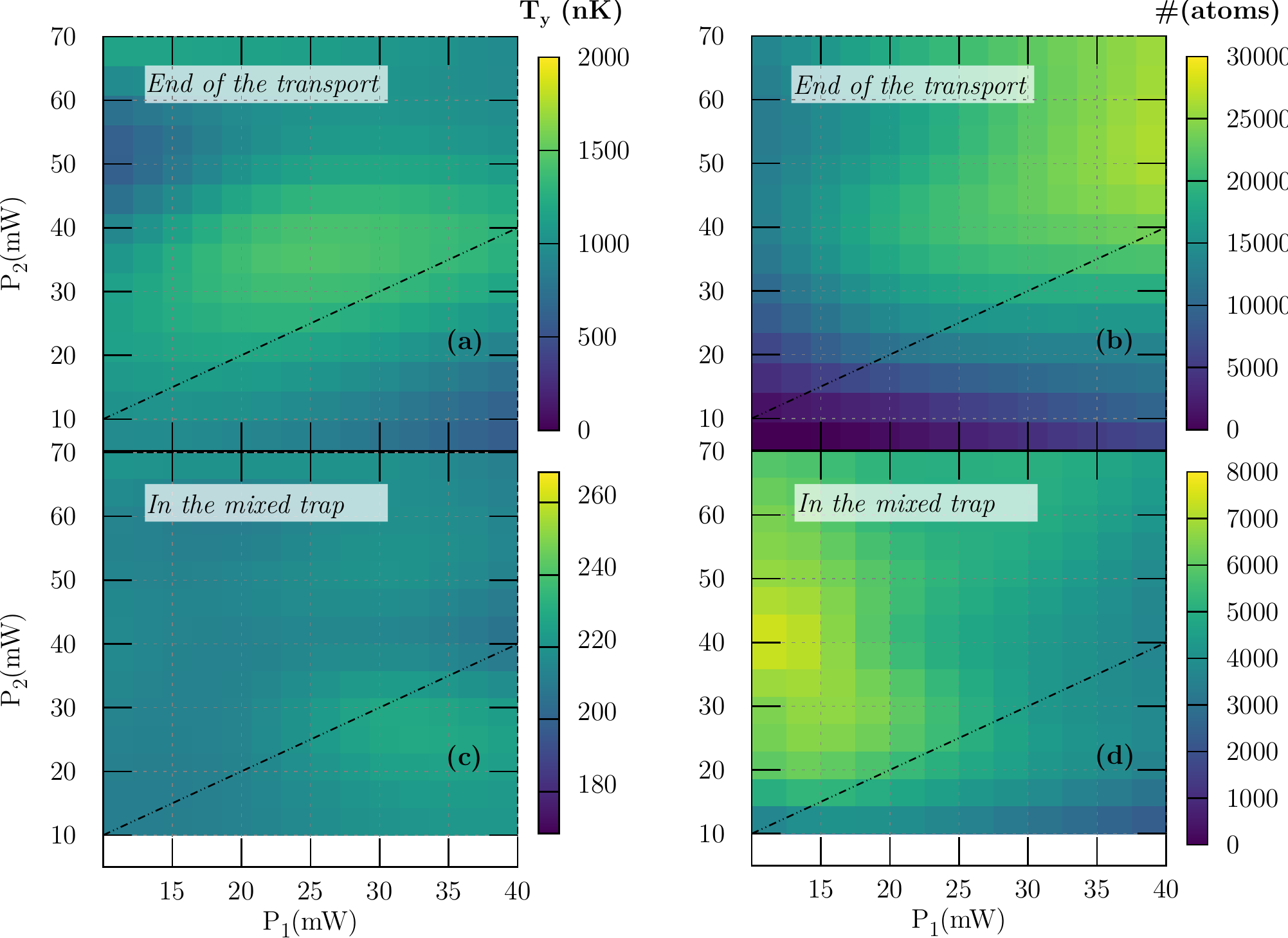}
  \end{center}
\caption{Influence of the powers in the Bloch beams on the number of atoms and vertical temperature $T_y$, either at the end of transport (a and b), or after their recapture in the mixed trap (during 1s) (c and d).}
\label{ResultsPGraphs}
\end{figure}

The influence of the powers in the Bloch beams during the transport is illustrated in figure \ref{ResultsPGraphs}-a. The larger the powers, the higher the depth, and the more efficient is the transport. However, for an efficient subsequent loading in the mixed trap, the power in the ``\textit{Bloch 1}" beam needs to be reduced, as shown in Figure \ref{ResultsPGraphs}-c, which displays the number of atoms in the mixed trap after a trapping time of 1s. We attribute this feature to the presence of residual parasitic static lattices, related to the reflection of the ``\textit{Bloch 1}" beam on the Raman and green lattice mirrors. 

Other parameters have also been varied without much improvement. We can for instance mention the influence of the depth used during the transport ($U_{ff}$). While decreasing the depth at this stage reduces in principle losses due to spontaneous emission, it comes at the price of a large decrease in the number of transported atoms, which we attribute to the reduction in the transverse confinement. Therefore this depth was kept constant, as high as during the acceleration and deceleration phases, for all the measurements presented here.

\begin{figure}[h]
  \begin{center}
    \includegraphics[scale=1]{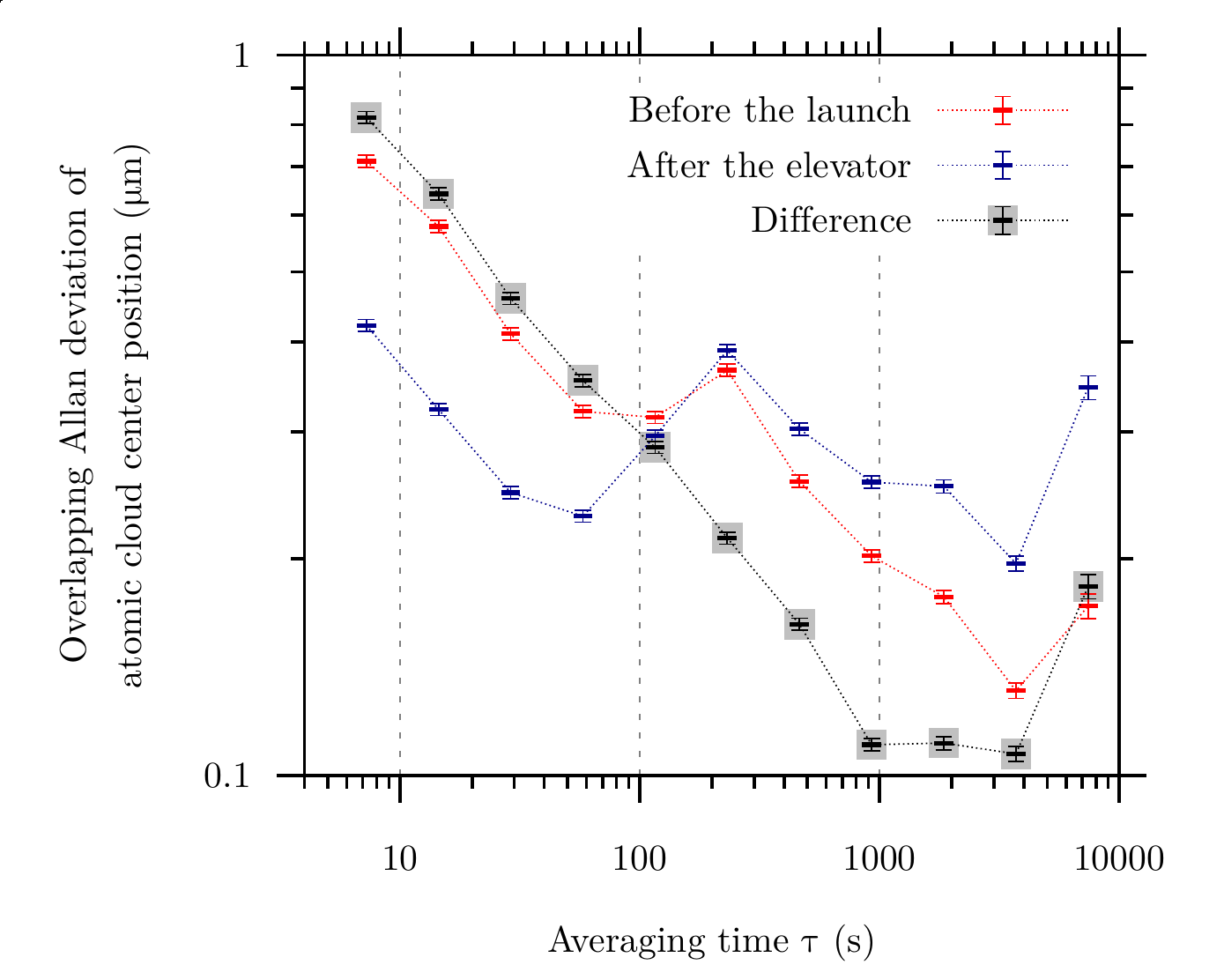}
  \end{center}
\caption{Overlapping Allan standard deviation of the center position of the atomic cloud before and after the transport}
\label{position_alla_dev}
\end{figure}

The position stability of the cloud after the transport is a key objective for our experimental setup. To check its evolution over time, we alternated absorption imaging measurements of the cloud before and after the transport. The corresponding Allan standard deviations of those two signals and of their difference are plotted in figure~\ref{position_alla_dev}.

A larger measurement noise is observed for the position before the launch in the bottom chamber, which we attribute to the larger size of the cloud, whose center position has been measured after a 6 ms thermal expansion in free fall, and to the larger readout noise of the bottom camera compared to the top one.

On both positions of the cloud, we observe a bump on the Allan deviation, occurring around 200s. This perturbation is correlated with the temperature cycles of our air conditioning system and is actually eliminated when calculating the difference of the two signals, since the Allan deviation of this difference averages as white noise down to about 100 nm at 1000s. This confirms that the travel distance is not affected by these thermal fluctuations, which are dominated by the fluctuations of the starting position of the atomic cloud, before the transport. 

This correlation between the top and bottom positions decreases over longer time scales, of hours or days, as can already be observed in figure~\ref{position_alla_dev} for timescales of a few thousand seconds. This could actually be due to deformations in the mechanical structure supporting our two imaging systems, rather than variations in the actual travel distance.

Finally, the final vertical size of the cloud will determine the spatial resolution of our force sensor. But, given the current resolution of our imaging system of about $9 \mu$m, we are actually not able to precisely determine this size, nor its fluctuations. Nevertheless, since the atoms remain trapped at all times, we expect the size of the atomic cloud to be identical to the size at the end of the evaporation (ie of order of a few micrometers in rms width).

\section{Conclusion}

We have investigated the efficiency of using Bloch oscillations to transport a cloud of ultracold atoms over a long-range distance of 30 cm, with an accurate control of their final position, size and temperature, and with the strong constraint of keeping them trapped at all times. We obtain a maximum efficiency of 25$\%$ for their transport, and in the end, we recapture about 10$\%$ of the initial number of atoms in the final trap, which combines a blue-detuned static lattice and a red-detuned progressive wave for transverse confinement. Spontaneous emission induces in our setup significant losses and heating, which could be reduced by operating at larger detunings, and thus intensities. At the end of the sequence, we are left with a cold and well localized sample of atoms, which we will use to probe short range interaction forces between the atoms and the dielectric mirror that retroreflects the lattice beam.

\section{Acknowledgments}

This research has been carried out in the frame of the QuantERA project TAIOL, funded by the European Union's Horizon 2020 Research and Innovation Programme and the Agence Nationale de la Recherche (ANR-18-QUAN-0015-01).

%%%%%%%%%%%%%%%%%%%%%%%%%%%%%%%%%%%%%%%%%%%%%%%%%%%%%%%%%%%%%%%%%%%%%%%%
%%%%%%%%%%%%%%%%%%%%%%%%%%%%%%%%%%%%%%%%%%%%%%%%%%%%%%%%%%%%%%%%%%%%%%%%
\pagebreak
\bibliographystyle{apsrev4-1}

\bibliography{ref}

\end{document}